\documentclass[nohyperref]{article}

\usepackage{microtype}
\usepackage{graphicx}
\usepackage{booktabs} 
\usepackage{amsmath}
\usepackage{bm}
\usepackage{multirow}
\usepackage{caption}
\usepackage{subcaption}
\usepackage{xcolor, soul}
\definecolor{light-gray}{gray}{0.85}  
\usepackage{hyperref}

\usepackage[accepted]{icml2022}

\usepackage{amsmath}
\usepackage{amssymb}
\usepackage{mathtools}
\usepackage{amsthm}

\usepackage[capitalize,noabbrev]{cleveref}

\theoremstyle{plain}

\theoremstyle{definition}

\theoremstyle{remark}

\usepackage[textsize=tiny]{todonotes}

\newcommand{\algname}{\textsc{ContentVec }}
\newcommand{\algnamens}{\textsc{ContentVec}}

\newcommand{\e}[1]{{\small $#1$}}

\icmltitlerunning{\textsc{ContentVec:} An Improved Self-Supervised Speech Representation by Disentangling Speakers}

\begin{document}

\twocolumn[
\icmltitle{\textsc{ContentVec:} An Improved Self-Supervised Speech Representation by Disentangling Speakers}



\icmlsetsymbol{equal}{*}

\begin{icmlauthorlist}
\icmlauthor{Kaizhi Qian}{equal,ibm}
\icmlauthor{Yang Zhang}{equal,ibm}
\icmlauthor{Heting Gao}{uiuc}
\icmlauthor{Junrui Ni}{uiuc}
\icmlauthor{Cheng-I Jeff Lai}{mit}
\icmlauthor{David Cox}{ibm}
\icmlauthor{Mark Hasegawa-Johnson}{uiuc}
\icmlauthor{Shiyu Chang}{ucsb}
\end{icmlauthorlist}

\icmlaffiliation{ibm}{MIT-IBM Watson AI Lab}
\icmlaffiliation{mit}{Massachusetts Institute of Technology}
\icmlaffiliation{uiuc}{University of Illinois at Urbana-Champaign}
\icmlaffiliation{ucsb}{University of California, Santa Barbara}

\icmlcorrespondingauthor{Kaizhi Qian}{kqian@ibm.com}
\icmlcorrespondingauthor{Yang Zhang}{yang.zhang2@ibm.com}

\icmlkeywords{Machine Learning, ICML}

\vskip 0.3in
]



\printAffiliationsAndNotice{\icmlEqualContribution} 

\begin{abstract}
Self-supervised learning (SSL) in speech involves training a speech representation network on a large-scale unannotated speech corpus, and then applying the learned representations to downstream tasks. Since the majority of the downstream tasks of SSL learning in speech largely focus on the content information in speech, the most desirable speech representations should be able to disentangle unwanted variations, such as speaker variations, from the content. However, disentangling speakers is very challenging, because removing the speaker information could easily result in a loss of content as well, and the damage of the latter usually far outweighs the benefit of the former. In this paper, we propose a new SSL method that can achieve speaker disentanglement without severe loss of content. Our approach is adapted from the HuBERT framework, and incorporates disentangling mechanisms to regularize both the teachers (masked prediction labels) and the students (learned representations). We evaluate the benefit of speaker disentanglement on a set of content-related downstream tasks, and observe a consistent and notable performance advantage of our speaker-disentangled representations.\footnote{Our code is available at \url{https://github.com/auspicious3000/contentvec}}
\end{abstract}

\section{Introduction}
Over the recent years, self-supervised learning (SSL) has emerged as a state-of-the-art solution to many speech processing problems with relatively few annotated data. The basic idea of SSL in speech is to train a speech representation network on large-scale \emph{unannotated} corpora, with an objective to capture and elicit meaningful speech structures and information. The resulting speech representation is then applied to the training of downstream tasks with a small amount of annotated data. Since the speech representation is already well-structured, it reduces the the dependency of downstream task training on large-scale datasets.

While speech SSL has demonstrated advantages in a surprisingly wide range of tasks, one of the primary foci of speech SSL is on tasks that process the \emph{content} of speech, such as speech recognition/phone classification, speech content generation, \emph{etc}. For these tasks, the most desirable speech representations should be the ones that can disentangle content information in speech from other interfering variations, such as speaker variations. However, among the most widely-used existing speech representations, few can achieve a reasonable disentanglement of speaker variations. For example, the \textsc{HuBERT} representation \citep{Hsu2021HuBERTSS} can achieve a speaker identification accuracy of up to 81.4\% on the SUPERB benchmark \cite{yang2021superb}. This observation suggests that there may still be room for performance gain for SSL on content-related speech processing tasks, if the disentanglement of speaker is adequately addressed.

However, it has been widely acknowledged that disentangling speakers is very challenging. Since no text annotations are accessible during the training of the speech representation network, any attempt to remove speaker variations from speech representation could easily lead to a loss of content information \cite{choi2021neural}. In most content-related downstream tasks, the cost of losing content information far outweighs the advantage in disentangling speakers. 


In this paper, we seek to investigate the following two research questions. First, is there a way to disentangle speaker variations during SSL training \emph{without} significant content loss? To this end, we propose \algnamens, an SSL framework that is adapted from the \textsc{HuBERT} training paradigm. The key idea of \textsc{HuBERT} is that by having some relatively poor speech representations, such as MFCC, serve as the teacher labels for the masked prediction task, one can derive speech representations (which are sometimes referred to as \emph{students}) that are far better than the teachers in many aspects, including content preservation. This inspires us that by combining \textsc{HuBERT}'s teacher-student framework with speaker disentanglement techniques, we could potentially restore the content loss caused by the latter. 

This has led us to the design of \algnamens, which incorporates into \textsc{HuBERT} three disentangling mechanisms - \emph{disentanglement in teachers} and \emph{disentanglement in students}, and \emph{speaker conditioning}. Specifically, disentanglement in teachers refers to removing the speaker information from the teacher labels. Disentanglement in students refers to introducing a regularization loss that directly enforces speaker invariance on the speech representations. Speaker conditioning refers to inputting speaker information to the masked prediction task, so that the need for the speech representation to encode speaker information is relieved. As we will show, all three modules are essential in shaping the speaker information flow across the speech representation network layers, and thereby achieve a superior disentanglement quality while keeping the content information intact.

The second research question we would like to explore is: How much performance gain, if any, can speaker disentanglment in SSL features contribute to downstream tasks? Our extensive evaluation shows that speaker disentanglement can achieve a consistent performance advantage over the baseline speech representations on content-related applications. 
The findings of this paper can shed some light on next-generation speech representations that can supply more targeted information to the downstream tasks and enable more powerful content processing directly on speech.

\section{Related Work}
\paragraph{Voice Conversion}
 Voice conversion is among the first research areas where speaker disentanglement is explored. The general trend follows the analysis-synthesis framework, where the analysis stage learns a speaker-independent speech representation that only preserves the content, and the synthesis stage uses the speaker-independent speech representation and the speaker-related variations to synthesize the conversion results.
 Much research focuses on learning better linguistic representations during the analysis stage and/or injecting speaker variations better during the synthesis stage. VAE-VC \cite{hsu2016voice} is an early attempt of directly using VAE for voice conversion. Afterward, \citet{Chou2018MultitargetVC} disentangles more speaker variations from the latent representation by discouraging the latent representation to be classified as the source speaker using an auxiliary speaker classifier on the latent representation. In contrast, ACVAE-VC \cite{kameoka2018acvae} indirectly encourages more speaker disentanglement by encouraging the conversion output to be correctly classified as the source speaker. Inspired by image style transfer, StarGAN-VC \cite{Kameoka2018StarGANVCNM}, StarGAN-VC2 \cite{kaneko2019stargan}, CycleGAN-VC \cite{Kaneko2018CycleGANVCNV}, and CycleGAN-VC2 \cite{Kaneko2019CycleganVC2IC} adapted StarGAN \cite{Choi2018StarGANUG} and CycleGAN \cite{Zhu2017UnpairedIT} respectively for voice conversion. AutoVC \cite{qian2019autovc} disentangles speakers and content by directly tuning the bottleneck dimensions of a vanilla autoencoder. The following AutoVC-F0 \cite{qian2020f0} improves pitch disentanglement by conditioning the synthesis stage on pitch representations. VoiceMixer \cite{Lee2021VoiceMixerAV} improves the content loss of AutoVC using similarity-based downsampling as the bottleneck. AdaIN-VC \cite{Chou2019OneshotVC} uses instance normalization to normalize out the speaker variations in the analysis stage, and AGAIN-VC \cite{Chen2021AgainVCAO} additionally uses an activation function to constrain the speaker variations from flowing into the synthesis stage. Instead of pursuing extreme speaker disentanglement, another slightly different track of research encourages the synthesis stage to use the supplied speaker variations by using partially disentangled content representations combined with speaker variations that are easier for the synthesis stage to utilize. SpeechSplit \cite{qian2020unsupervised}, AutoPST \cite{Qian2021GlobalPS}, and NANSY \cite{choi2021neural} perturb the speaker variations during the analysis stage to encourage the synthesis stage to use the supplied more stable speaker representations. In particular, \citet{polyak2021speech} and NANSY start with the self-supervised speech representations as the partially disentangled content representation.
\vspace{-0.2in}
\paragraph{Self-supervised Learning in Speech}
Learning self-supervised speech representation usually encodes the speech feature into context representations followed by pretext tasks to extract content information, which mainly has two tracks. The first track is generative learning. \citet{Chung2019AnUA,Chung2020GenerativePF} uses Autoregresstive Predictive Coding (APC) for self-supervised representation learning. Mockingjay \cite{Liu2020MockingjayUS} learns speech representation by predicting the current frame given both the past and future contexts. TERA \cite{Liu2021TERASL} learns speech representation by reconstructing acoustic
frames from their altered counterparts. DeCoAR 2.0 \cite{Ling2020DeCoAR2D} reconstructs the frames from their vector-quantized counterparts. \citet{Wang2020UnsupervisedPO} reconstructs masked frames. The second track is discriminative. \citet{Oord2018RepresentationLW} uses constrastive predictive coding to learn multi-modal represenations including speech. Wav2vec \cite{Schneider2019wav2vecUP} learns to predict the future samples from distractors. Wav2vec 2.0 \cite{Baevski2020wav2vec2A}, an end-to-end version of vq-wav2vec \cite{Baevski2020vqwav2vecSL}, learns to identify the true vq-quantized frame among the distractors using contrastive loss. \citet{Kharitonov2021DataAC} significantly improves CPC-based SSL with speech data augmentation. \citet{Zhang2020PushingTL} pushes the limits of SSL using noisy student training by giant Conformer models pre-trained using wav2vec 2.0. Hubert \cite{Hsu2021HuBERTSS} predicts masked frames pre-quantized using k-means. ILS-SSL \cite{Wang2021SelfSupervisedLF} further improves Hubert by adding masked prediction loss on intermediate layers. Besides, there are also research using multiple tasks \cite{Pascual2019LearningPS,Ravanelli2020MultiTaskSL,Chung2021W2vBERTCC} or using both labeled and unlabeled data \citep{Wang2020UnsupervisedPO}.

\section{Approach}
\begin{figure*}
    \centering
    \includegraphics[width=0.85\linewidth]{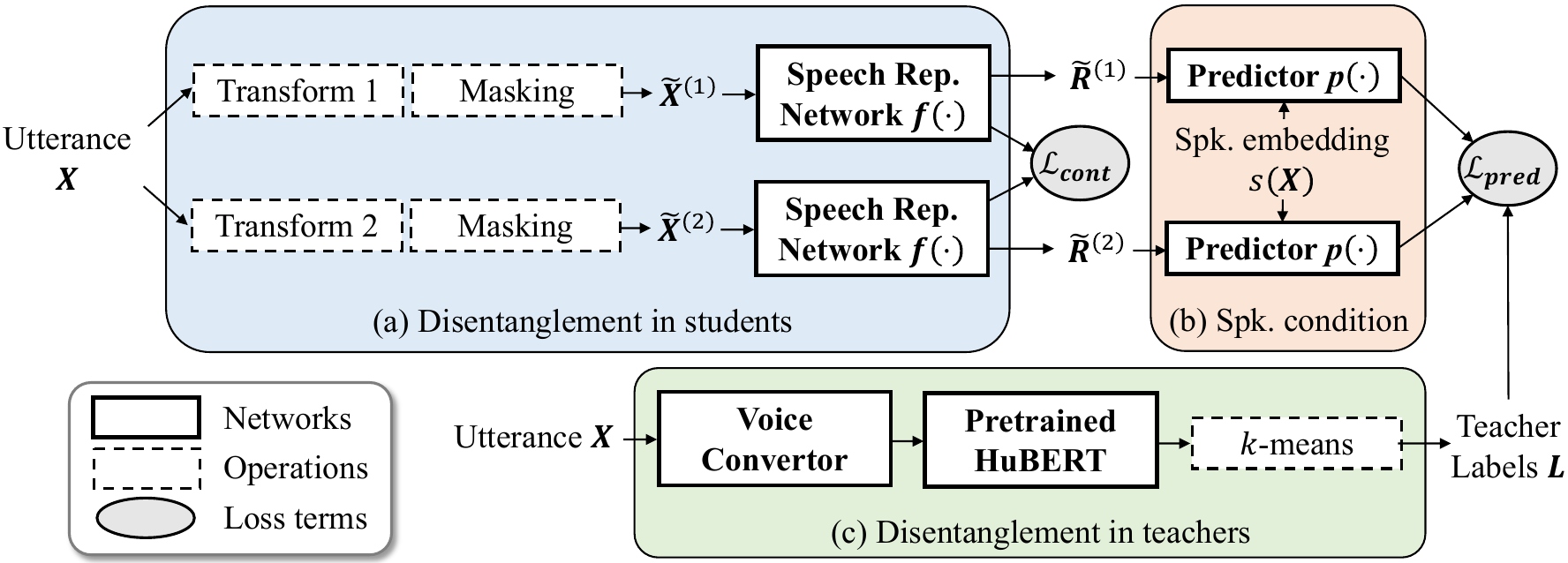}
    \caption{The overall structure of \algnamens. Each colored block represents one disentanglement module in \algnamens.}
    \label{fig:structure}
\end{figure*}

In this section, we will introduce of our approach. We use upper-cased latters, \e{X} and \e{\bm X}, to represent random scalars and vectors, respectively, and lower-cased latters, \e{x} and \e{\bm x}, to represent deterministic scalars and vectors, respectively.

\subsection{Problem Formulation}

Denote \e{\bm X = [\bm X_1, \cdots, \bm X_T]} as the sequence of a speech features, where \e{\bm X_t} is the speech feature vector at frame $t$, and $T$ is the total number of frames. Our goal is to learn a speech representation network \e{\bm R = f(\bm X)}, where \e{\bm R = [\bm R_1, \cdots, \bm R_T]} and \e{\bm R_t} is the representation for frame $t$. \e{\bm R} should desirably satisfy the following two properties.
\vspace{-3mm}
\begin{itemize}
    \item  \e{\bm R} should preserve as much content information as possible, and the content information roughly corresponds to the phonetic/text transcriptions of the utterance.
    \vspace{-1mm}
    \item \e{\bm R} should be invariant across speaker variations.
\end{itemize}
As mentioned, the pursuit of one goal can easily compromise another. In the following , we will describe our method to strike a better balance and discuss the rationale behind.

\subsection{The General Framework}

The \algname framework builds upon the mask-prediction framework of \textsc{HuBERT}. Specifically, there are three components in the \textsc{HuBERT} framework: 1) the speech representation network $f(\cdot)$, 2) the predictor $p(\cdot)$, and 3) the teacher label generator $g(\cdot)$.

During training, the speech representation network takes the partially masked speech utterance, \e{\tilde{\bm X}}, as the input, and produces a representation for the masked speech sequence, \e{\tilde{\bm R} = f(\tilde{\bm X})}. On the other hand, the teacher label generator generates a label sequence \e{\bm L = g(\bm X)} from the \emph{unmasked} speech. The goal of the predictor is to predict the teacher labels \e{\bm L} from the masked speech representation \e{\tilde{\bm R}}. The teacher label generator $g(\cdot)$ is usually predefined and fixed during training. The other two modules, $f(\cdot)$ and $p(\cdot)$, are trained jointly to minimize the following prediction loss:
\begin{equation}
    \small
    \mathcal{L}_{pred}= \mathbb{E}[\ell_m(p \circ f(\bm \tilde{\bm X}),  g(\bm X))],
    \label{eq:L_pred}
\end{equation}
where \e{\ell_m} denotes the cross-entropy loss computed over the \emph{masked} frames only. To make our description more intuitive, we will refer to \e{f(\tilde{\bm X})} as \emph{students}, and \e{g(\bm X)} as \emph{teachers}.

It has been reported \citep{Hsu2021HuBERTSS} that even if the \textsc{HuBERT} teacher is poor (\emph{e.g.}, losing content), the student can still preserve the content far better than the teacher, thanks to the masked prediction mechanism. This observation inspires us to test the hypothesis that one can combine speaker disentanglement techniques (potentially causing loss of content) with the masked prediction framework, and in this way, preserve content more faithfully than 
using a speaker disentanglement algorithm on its own. Since teachers, students, and the predictor are three major components of the masked prediction, \algname introduces three disentanglement mechanisms, \emph{disentanglement in teachers}, \emph{disentanglement in students}, and \emph{speaker conditioning}, to tackle the three components respectively, as shown in Figure~\ref{fig:structure}.

\subsection{Disentanglement in Teachers}

Disentanglement in teachers aims to remove the speaker information in the teacher labels. Recently, there has been marked progress in unsupervised voice conversion systems, which can now significantly obscure the source speaker information without losing too much content \cite{polyak2021speech}. Inspired by this, we adopt a voice conversion model to convert all utterances to the same speaker before generating the teacher labels.

Specifically, as shown in Figure~\ref{fig:structure}(c), the teacher labels, \e{\bm L = g(\bm X)}, are generated via the following three steps. First, all the utterances \e{\bm X} in the training set are converted to a single speaker using a competent unsupervised voice conversion system.
Second, the converted utterances are passed through a pre-trained unsupervised speech representation network, in our case \textsc{HuBERT}, to generate a set of speech representations, which should contain very little speaker information. Finally, the speech representations are quantized to discrete teacher labels using $k$-means clustering.

It is worth noting that although the teacher speech representation described above already achieves speaker disentanglement, its content preservation is not satisfactory because any voice conversion systems sometimes (for some speakers) cause a non-negligible content loss \cite{choi2021neural}. In order to ameliorate this shortcoming of modern voice conversion, we use voice conversion as a teacher to train better students, instead of directly applying its output to downstream tasks.

\subsection{Disentanglement in Students}

Disentanglement in students enforces speaker-invariant student representations, which can be achieved with \textsc{SimCLR} \cite{chen2020simple}, a contrastive-learning-based algorithm.

Specifically, as shown in Figure~\ref{fig:structure}(a), each speech utterance, \e{\bm X}, is passed into two random transformations that alter only the speaker information, before it is masked. Denote the two masked, transformed copies of \e{\bm X} as \e{\tilde{\bm X}^{(1)}} and \e{\tilde{\bm X}^{(2)}}. Then, this pair of utterances are passed through the speech represetnation network, $f(\cdot)$, to generate the representations \e{\bm R^{(1)}} and \e{\bm R^{(2)}}, and the following contrastive loss is introduced to penalize dissimilarity between \e{\bm R^{(1)}} and \e{\bm R^{(2)}}:
\begin{equation}
    \small
    \begin{aligned}
    \mathcal{L}_{contr} = & \sum_{t=1}^T \frac{\exp(\textrm{cossim}(\bm R^{(1)}_t, \bm R^{(2)}_t) / k)}{\sum_{\tau \in \{t\} \cup \mathcal{I}_t} \exp(\textrm{cossim}(\bm R^{(1)}_t, \bm R^{(1)}_\tau)/k)} \\
    + & \sum_{t=1}^T \frac{\exp(\textrm{cossim}(\bm R^{(2)}_t, \bm R^{(1)}_t)/k)}{\sum_{\tau \in \{t\} \cup \mathcal{I}_t} \exp(\textrm{cossim}(\bm R^{(2)}_t, \bm R^{(2)}_\tau)/k)},
    \end{aligned}
    \label{eq:L_contr}
\end{equation}
where cossim$(\cdot, \cdot)$ denotes the cosine similarity, and \e{\mathcal{I}_t} denotes a set of random time indices at which the representations are chosen as the negative examples for time $t$. The contrastive loss consists of two terms so that it is symmetric with respect to \e{\bm R^{(1)}} and \e{\bm R^{(2)}}. According to Equation~\eqref{eq:L_contr}, the negative examples for the utterance pair, \e{(\bm R_t^{(1)}, \bm R_t^{(1)})}, are uniformly randomly drawn from the remaining frames within the \emph{same} utterances. As an extention to Equation~\eqref{eq:L_contr}, the contrastive loss can be applied to an intermediate layer, instead of the final layer, of $f(\cdot)$. Section~\ref{subsec:flow} will discuss how the choice of layer in which the contrastive loss is imposed would affect the disentanglement behavior.

The biggest challenge of applying the contrastive loss is how to design a random transformation that only alters the speaker identity of the utterance with minimal changes in the other aspects. To this end, we adopt the random transformation algorithm proposed by \citet{choi2021neural}. Specifically, the algorithm consists of three steps of transformations. First, all the formant frequencies within an utterance are scaled by a factor of $\rho_1$; second, F0 in every frame is scaled by a factor of $\rho_2$; finally, a random equalizer is applied to accommodate any channel effects. $\rho_1$ and $\rho_2$ are both randomly drawn from the uniform distribution \e{\mathcal{U}([1, 1.4])}, and then flipped to their reciprocals with probability $0.5$. Since the majority of voice information resides in the formant frequency and F0 frequency ranges (e.g.,~\cite{eide1996parametric}), while content information resides in the relative formant frequency ratios~\cite{stevens1987relational}, uniform scaling of all the formant and F0 tends to change the speaker information while retaining the content.

To further strengthen the invariance, the same random transformations are also applied to the student representations in the masked prediction task, \emph{i.e.}, Equation~\eqref{eq:L_pred} is modified as
\begin{equation}
    \small
    \mathcal{L}_{pred} = \mathbb{E}[\ell_m(p \circ f(\bm \tilde{\bm X}^{(1)}),  g(\bm X)) + \ell_m(p \circ f(\bm \tilde{\bm X}^{(2)}),  g(\bm X))].
\end{equation}
Again, the masked prediction loss is applied to both \e{f(\tilde{\bm X}^{(1)})} and \e{f(\tilde{\bm X}^{(2)})} for symmetry.

\subsection{Speaker Conditioning}

Although disentanglement in teacher can remove the majority of the speaker information from the teacher labels, certain speaker information would remain. As a result, the student representations are undesirably forced to carry the same amount of speaker information as the teachers do in order to reasonably predict the teacher labels. To break this entailment between the speaker information in students and in teachers, we feed the speaker embeddings to the predictor. Speaker embeddings are produced by a speaker embedding network, in our case a pre-trained GE2E \citep{wan2018generalized}, which takes a speech utterance as input and outputs a vector summarizing the speaker information in the utterance. Therefore, by conditioning the predictor on the speaker embedding, we can supply whatever speaker information is needed for the mask prediction task, so that the students do not have to carry the speaker information themselves. 

Formally, the masked prediction loss now becomes
\begin{equation}
    \small
    \begin{aligned}
    \mathcal{L}_{pred} = & \mathbb{E}[\ell_m(p( f(\bm \tilde{\bm X}_1), s(\bm X)),  g(\bm X)) \\
    & + \ell_m(p( f(\bm \tilde{\bm X}_2), s(\bm X)),  g(\bm X))],
    \end{aligned}
\end{equation}
where \e{s(\bm X)} denotes the speaker embeddings. The final loss is the superposition of the prediction and contrastive losses:
\begin{equation}
    \small
    \mathcal{L} = \mathcal{L}_{pred} + \lambda \mathcal{L}_{contr}.
    \label{eq:loss_combine}
\end{equation}

As can be observed, although \algname requires speaker labels to identify speaker information, speaker labels are only used in pre-training the speaker embedding network. The training of \algname itself only requires speaker embeddings, not speaker labels. Since the speaker embedding network is pre-trained on a separate dataset, and can well generalize to unseen speakers, the training set for \algname does not need to contain any speaker labels.

\subsection{An Information Flow Perspective}
\label{subsec:flow}

\begin{figure}
    \centering
    \includegraphics[width=0.8\columnwidth]{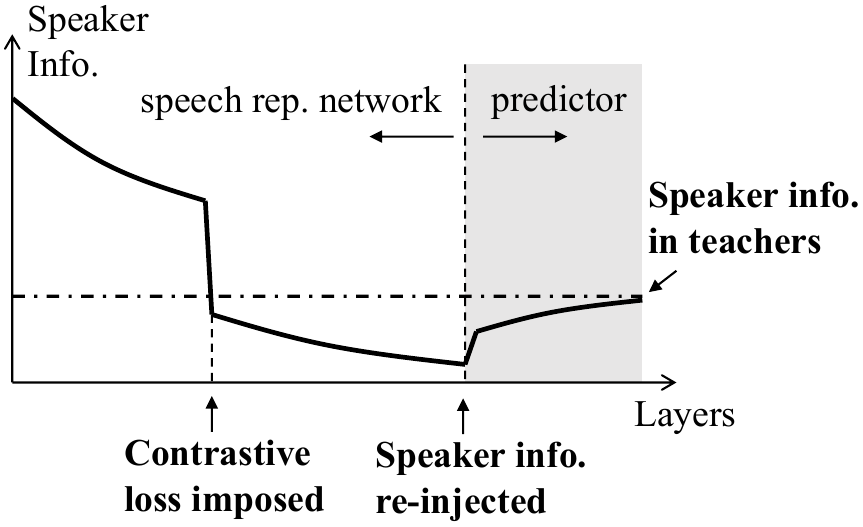}
    \caption{A conceptual curve of how speaker information changes through the network layers.}
    \label{fig:concept}
\end{figure}

To provide an intuitive illustration of how the aforementioned modules work collaboratively towards disentangling speakers, Figure~\ref{fig:concept} shows a conceptual curve of how the amount of speaker information changes along different layers of the speech representation network $f(\cdot)$ and the predictor $p(\cdot)$. The vertical axis denotes the amount of speaker information, and the horizontal axis denotes the number of layers. The white area denotes the speech representation network layers, and the grey area denotes the prediction layers, which are on top of the speech representation network. To the left, the speaker information is equal to the full speaker information in the input utterance. To the right, the speaker information should be roughly equal to the speaker information in the teacher labels, which is much lower than that in the input but is still not zero. Due to the information processing inequality, the speaker information is monotonically decreasing as the layer progresses, except for the predictor layers where speaker information is re-injected.

As can be observed, there are two places where the speaker information undergoes abrupt changes. The first is where the contrastive loss (Equation~\eqref{eq:L_contr}) is imposed, and the speaker information is largely reduced. The second is where the speaker information is re-injected, and the speaker information slightly increases. As a result, the speaker information should reach its minimum at the intersection between the speech representation network and the predictor. Figure~\ref{fig:concept} shows that all the modules in \algname are essential to a successful speaker disentanglement.


\section{Experiments}
In this section, we will evaluate \algname on an extensive set of content-related tasks. In particular, we would like to investigate whether disentangling speakers has benefit in real-world tasks, and how large the benefit would be. Further experimental details can be found in Appendix B.

\begin{table*}[t]
\caption{Results on zero-resource content probe and language modeling.}
\label{tab:zero-speech}
\begin{center}
\begin{small}
\begin{tabular}{lccccccr}
\toprule
Model & ABX(w)	$\downarrow$ & ABX(a) $\downarrow$ & Lexical$\downarrow$ & Syntactic $\downarrow$ & PPX $\downarrow$ & VERT	$\downarrow$ & AUC$\downarrow$ \\
\midrule
\algnamens      & \bf 5.13 & \bf 6.32 & \bf 33.27 & \bf 43.95 & \bf 650.04    & \bf 46.05     & \bf 45.01 \\
\textsc{HuBERT-iter} & 6.01     & 7.20     & 34.00     & 44.36     & 739.12     & 47.55     & 53.28 \\
\textsc{HuBERT}      & 6.06     & 7.37     & 36.19     & 46.48     & 790.17     & 54.35     & 75.23 \\
\textsc{Wav2Vec 2.0}     & 8.70     & 10.34    & 35.93     & 46.40     & 840.34     & 58.59     & 88.83  \\
\bottomrule
\end{tabular}
\end{small}
\end{center}
\vskip -0.1in
\end{table*}

\subsection{Configurations}
\label{subsec:config}

\paragraph{Implementation Details} The speech representation network of \algname has the same architecture as \textsc{HuBERT}. According to Section~\ref{subsec:flow}, the speaker-disentanglement is optimal at the output of the speech representation network, so we select the output-layer representation as the \algname features. The contrastive loss is imposed at the last but five layer, the temperature $k$ is set to 0.1, and the contrastive loss weight $\lambda=\text{1e-5} * num\_train\_steps$, which linearly increases to 10 when training for 100k steps. A parameter sensitivity evaluation will be provided in Section~\ref{subsec:ablation}.

The predictor of \algname contains three transformer layers \emph{without} layer drop. The frame masking scheme and prediction logit generation for the masked prediction task are the same as \textsc{HuBERT}.

To generate the teacher labels, we use the voice converter proposed by \citet{polyak2021speech}.
We re-train the voice converter on a subset of 200 speakers chosen from the \texttt{Librispeech} dataset \cite{Panayotov2015LibrispeechAA}, using the publicly-released \textsc{HuBERT} base model\footnote{\url{https://github.com/pytorch/fairseq/tree/main/examples/hubert}} with 100 clusters. 
The model checkpoint that gives the highest average target speaker similarity is selected.
The teacher utterances are then generated by converting the entire \texttt{Librispeech-960h} to the voice of that selected target speaker with the selected model. After passing the converted utterance to the pretrained \textsc{HuBERT}, the seventh layer feature representation is chosen, because compared with the commonly chosen sixth, the seventh layer has a lower speaker classification accuracy and yet comparable teacher quality \cite{Hsu2021HuBERTSS}. The number of clusters for quantizing final teacher labels is 100. 

\paragraph{Baselines and Dataset} The following baselines are included in the evaluation.

$\bullet$ \textsc{Wav2Vec 2.0} \cite{Baevski2020wav2vec2A}: Following \citet{Lakhotia2021OnGS}, the 14th layer representation is chosen.

$\bullet$ \textsc{HuBERT} \cite{Hsu2021HuBERTSS}: We adopt the publicly-released pretrained model in \citet{Ott2019fairseqAF}. Following \citet{Hsu2021HuBERTSS}, the sixth layer representation is chosen.

$\bullet$ \textsc{HuBERT-iter}: Since \algname is guided by a pretrained \textsc{HuBERT} as teachers, for a fair comparison, we introduce another \textsc{HuBERT} baseline that is trained by the same pretrained \textsc{HuBERT} as teachers (except that no voice conversion is performed). This baseline controls the benefit of iterative training. We performed the same teacher quality evaluation as in \citet{Hsu2021HuBERTSS}, and identified layer eight as the best performing layer.

\algname and all the baselines are trained on the \texttt{Librispeech} dataset \citep{Panayotov2015LibrispeechAA}. If the evaluation task requires discrete representations, all the representations will be quantized to 100 clusters by $k$-means. Otherwise, the continuous representations will be used.

\subsection{Zero-shot Content Probe}
\label{subsec:zero-speech}

The first set of experiments we would like to evaluate is the set of zero-shot probing tasks proposed in the Zero-Resource Speech Challenges \cite{Rivire2021TowardsUL, Dunbar2021TheZR}, because they require a high alignment between the discrete representation and phonetic content. For some of these tasks, a language model trained on the discrete speech representations is needed. We use the same language model and hyperparameter setting as in \citet{Lakhotia2021OnGS}, which is the \emph{transformer LM big architecture} implemented in \texttt{fairseq} \citep{Ott2019fairseqAF}. 
We evaluate on four tasks.

$\bullet$ \textbf{ABX(w):} Given a pair of words with one difference in phoneme and a test word containing the same phoneme as one of the two words, ABX measures the probability that the test phoneme representation is closer to the representation of the correct phoneme in the word pair than to that of the incorrect phoneme. `(w)' indicates that the comparison is `within' the same speaker.

$\bullet$ \textbf{ABX(a):} Same as ABX(w), except that the test utterance is uttered by a different speaker. `(a)' indicates that the comparison is `across' different speakers.

$\bullet$ \textbf{Spot the Word (Lexical):} Spot the word measures the accuracy of identifying the correct word from a pair of real/fake words, based on the perplexity of the language model.

$\bullet$ \textbf{Acceptability Judgment (Syntactic):} Acceptability judgement measures the accuracy of identifying the syntactically correct sentence from a pair of correct/incorrect sentences, based on the perplexity of the language model.

Table~\ref{tab:zero-speech} (left four columns) shows the results of the zero-shot probing tasks. There are three key observations. First, \algname achieves consistent advantage on all four metrics, demonstrating that speaker disentanglement does help in these tasks. However, the size of the performance gain varies across different tasks and across different number of clusters. The benefit is largest for the phonetic-level tasks, ABX(w) and ABX(a). On the other hand, for lexical- and semantic-level tasks, the benefit is smaller. We believe that this is because the performance in these tasks not only depends on the quality of the speech representation, but also on the language model. Our second observation is that \textsc{HuBERT-iter} consistently outperforms \textsc{HuBERT}, which confirms that there is a benefit in iterative training, and this is why it is very important to include \textsc{HuBERT-iter} as a baseline for a fair comparison. Finally, note that there is a slight difference between our results for \textsc{HuBERT} and the results reported in \citet{Lakhotia2021OnGS}. This is likely because the publicly-released model that we use is different from the model used in \citet{Lakhotia2021OnGS}, in terms of number of clusters in teachers, batch size, the number of GPUs \emph{etc}. However, since \algname and \textsc{HuBERT-iter} are both derived from \textsc{HuBERT}, we expect a similar performance gap among these three methods if a different \textsc{HuBERT} model is used.

\begin{table*}[t]
\caption{Results on SUPERB tasks within ``Content'' and ``'Semantics'' categories.} 
\label{tab:superb}
\begin{center}
\begin{small}
\begin{tabular}{l|c|c|c|c|c|cc}
\toprule
 Tasks & PR & ASR & KS & QbE & IC & \multicolumn{2}{c}{SF} \\
        Metrics & PER $\downarrow$ & WER $\downarrow$ & ACC $\uparrow$ & MTWV $\uparrow$ & ACC $\uparrow$ & F1 $\uparrow$ & CER $\downarrow$ \\
        \midrule
        \textsc{ContentVec}      & \textbf{0.049}     & \textbf{5.7}       & \textbf{0.964}      & 0.0590     & \textbf{0.991}      & \textbf{0.896}     & \textbf{0.236} \\
        \textsc{HuBERT-iter}      & 0.052     & 6.5       & 0.963      & \textbf{0.0891}     & 0.983      & 0.886     & 0.259 \\
        \textsc{HuBERT}      & 0.054     & 6.4       & 0.963      & 0.0736     & 0.983      & 0.885     & 0.256 \\
        \bottomrule
        \end{tabular}
\end{small}
\end{center}
\end{table*}

\subsection{Language Modeling}
\label{subsec:language_model}

Language models built directly on the discrete speech representations can be applied to many content-related speech generation and analysis tasks. We would like to explore whether a disentangled speech representation can contribute to a language model with higher quality. To this end, we use the same language models as in Section~\ref{subsec:zero-speech} to generate random speech representation sequences under different temperatures, and resynthesize speech from these sequences using \textsc{Tacotron} as in \citet{Lakhotia2021OnGS}. For each temperature level, we compute the perplexity (PPX) and variety (VERT) score of the transcript as proposed in \citet{Lakhotia2021OnGS}. All the PPX-VERT pairs at different temperature levels form a curve depicting the quality-variety trade-off of each language model. We report the following three metrics.

$\bullet$ \textbf{PPX at oracle VERT:} The perplexity score when the VERT score equals the VERT score of true text.

$\bullet$ \textbf{VERT at oracle PPX:} The VERT score when the PPX score equals the PPX score of the true text.

$\bullet$ \textbf{AUC:} Area under the perplexity-VERT curve.


Table~\ref{tab:zero-speech} (right columns) shows the results. 
As can be seen, \algname achieves a significantly lower PPX score than all the baselines and a slightly lower VERT score. This indicates that the improvement in speaker disentanglement contributes to the correctness of the language model. This observation shows that speaker disentanglement significantly helps in improving the speech generation quality. Some qualitative results are shown in Appendix~\ref{app:generation}.

\subsection{SUPERB Experiments}

To extend our evaluation to supervised tasks, we use SUPERB \citep{yang2021superb}, a benchmark dataset containing an extensive list of supervised speech processing tasks. We select the subset of tasks belonging to the categories of ``content'' and ``semantic.'' These tasks include phone recognition (PR), automatic speech recognition (ASR), keyword spotting (KS), Query by Example Spoken Term Detection (QbE), intent classification (IC), and slot filling (SF). Detailed descriptions of these tasks can be found in \citet{yang2021superb}. During the training of these tasks, the speech representation networks are frozen.

Unlike the tasks discussed in the previous sections, the SUPERB tasks use the \emph{continuous} representations rather than the discrete ones. Therefore, the \textsc{HuBERT} baseline, which is trained with a 500-class teacher and prolonged training iterations, is expected to provide more information than do \algname and \textsc{HuBERT-iter}, which are both trained with a 100-class teacher. Therefore, we retrain \textsc{HuBERT} and \textsc{HuBERT-iter} with matched number of teacher clusters and training iterations for a fair comparison. Table~\ref{tab:superb} lists the results on the SUPERB tasks. We can observe that \algname generally outperforms both \textsc{HuBERT-iter} and \textsc{HuBERT}. This observation verifies that the benefit of speech disentanglement can generalize to content-related supervised tasks.

\subsection{Speaker \& Accent Classification}
\label{subsec:sid}

\begin{figure}
    \centering
    \includegraphics[width=0.8\columnwidth]{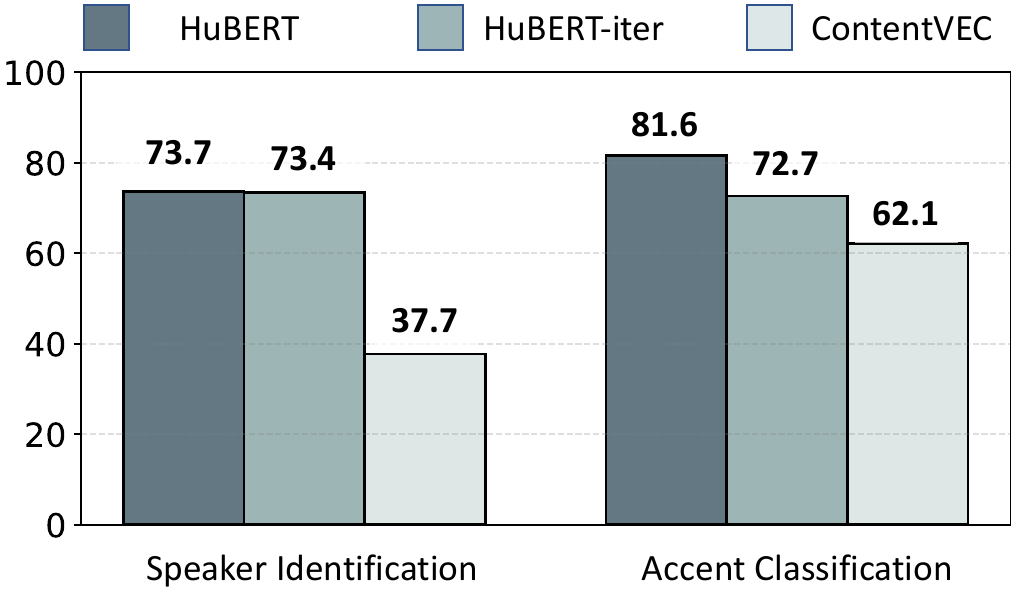}
    \caption{Bar plots of SID accuracy (left) and accent classification accuracy (right)}
    \label{fig:sid_acc}
    \vspace{-0.2in}
\end{figure}

Speaker identification serves as a proxy for speaker disentanglement. In addition, we are also interested in whether our speaker disentanglement algorithm removes regional accents or not: is regional accent primarily communicated by phoneme content, or by speaker-dependent style? We therefore evaluate the quality of speaker disentanglement on the speaker identification task (SID) in the SUPERB benchmark, and an accent classification task using the \texttt{L2-ARCTIC} dataset, which contains 7 accent groups.

Figure~\ref{fig:sid_acc} plots the accuracy of the two classification tasks. As can be seen, \algname sharply reduces the accuracy in both tasks. In the SID task, the reduction is as high as 36\% compared to the \textsc{HuBERT-iter}, indicating that the speaker disentanglement mechanisms in \algname are very effective. \textsc{HuBERT-iter} has a slightly lower SID accuracy than \textsc{HuBERT}, which shows that iterative training reduces the amount of speaker information. In the accent classification task, the reduction is also significant, which verifies that disentangling speaker information also reduces accent information to some degree.

\subsection{Voice Conversion}

The existing mainstream voice conversion systems follow an encoder-decoder paradigm, where the encoder derives a speech representation with speaker information disentangled and the decoder synthesizes the speech signal conditional on speaker embedding/labels. \citet{polyak2021speech} further shows that a decoder built directly on top of self-supervised speech representations suffices to produce state-of-the-art voice conversion results. To evaluate whether the advantage of speaker disentanglement in \algname translate to better voice conversion, we run the voice conversion model in \citet{polyak2021speech} on \algname and other baseline speech representations and evaluate the speaker similarity of the converted speech to the target speaker. Instead of using discrecitized speech representations as in \citet{polyak2021speech}, we consider a much more challenging setting where the continuous speech representations are used.
The models are trained and evaluated on \texttt{Librispeech}. Our test set contains the same speakers as the training set. To reduce the evaluation complexity, while the source speakers are from the entire test set, the target speakers are from a subset of 20 speakers, 10 are from the `clean' subset and 10 are from the `other' subset. Our evaluation is thus divided into four scenarios, clean to clean (C2C), clean to other (C2O), other to clean (O2C), and other to other (O2O).

Table~\ref{tab:vc} shows the average cosine similarity of the d-vectors \citep{Heigold2016EndtoendTS} between the converted speech and the target speakers. As can be observed, thanks to the inherent speaker disentanglement capability of the neural decoder, the \textsc{HuBERT}-based voice conversion model already has a decent speaker similarity, which is consistent with the findings by \citet{polyak2021speech}. However, \algname is able to further advance the performance by a significant margin, further verifying the advantage of \algnamens's speaker disentanglement quality. We also observe that \textsc{HuBERT-iter}, which has a slightly better speaker disentanglement property according to Section~\ref{subsec:sid}, also improves over \textsc{HuBERT} in terms of speaker similarity in this task.

\begin{table}[t]
\caption{Average cosine similarity (\e{\uparrow}) of the d-vectors between the converted speech based on different speech representations and the target speakers.}
\label{tab:vc}
\begin{center}
\begin{small}
\begin{tabular}{l|c|c|c|c}
\toprule
Setting & C2C & O2C & C2O & O2O \\
        \midrule
        \textsc{ContentVec}  & \textbf{0.9316} & \textbf{0.9277} & \textbf{0.9150} & \textbf{0.9257}    \\
        \textsc{HuBERT-iter} & 0.9286 & 0.9243 & 0.9050 & 0.9215  \\
        \textsc{HuBERT}      & 0.9029 & 0.8982 & 0.8848 & 0.9036 \\
        \bottomrule
        \end{tabular}
\end{small}
\end{center}
\end{table}

\subsection{Ablation Studies}
\label{subsec:ablation}

\begin{table*}[t]
\caption{Results of ablation studies.}
\label{tab:ablation}
\begin{center}
\begin{small}
\begin{tabular}{l|c|ccc|ccc|cccc}
\toprule
&& \multicolumn{3}{|c|}{Removing each module}  & \multicolumn{3}{|c|}{Position of $\mathcal{L}_{cont}$} & \multicolumn{4}{|c}{Weight of $\mathcal{L}_{cont}$ ($\lambda$)} \\
 & \algnamens & \textsc{No-DT} & \textsc{No-DS} & \textsc{No-Cond} & \textsc{N1} & \textsc{N4} & \textsc{N8} & 1e-6 & 5e-6 & 2e-5 & 5e-5\\
\midrule
ABX(w) $\downarrow$   & 5.13 & 5.67 & 5.62 & 5.76 & 5.76 & 5.19 & 5.14 & 5.18 & 5.10 & 5.18 & 5.22\\
ABX(a) $\downarrow$    & 6.32 & 6.92 & 6.93 & 7.08 & 7.03 & 6.43 & 6.40 & 6.36 & 6.39 & 6.51 & 6.58 \\
PNMI $\uparrow$            & 0.590 & 0.576 & 0.577 & 0.593 & 0.545 & 0.587 &  0.584 & 0.586 & 0.586 & 0.584 & 0.582 \\
\bottomrule
\end{tabular}
\end{small}
\end{center}
\vskip -0.1in
\end{table*}

This section evaluates different variants of \algname to understanding the contribution of each model design choice. Since there are a large number of models, we only select the following three efficient yet representative metrics: the phone normalized mutual information between the discrete representations and ground truth phonetic unit (PNMI) proposed in \citet{Hsu2021HuBERTSS}, ABX(w) accuracy, and ABX(a) accuracy. 
For each model variant, we report the best-layer results under each respective metric.

\paragraph{Contribution of Each Disentanglement Module} To measure how much each of our three major disentanglement mechanisms (disentanglement in teachers, disentanglement in students, and speaker conditioning of the predictor) contribute to the overall performance, we build variants of \algname with each individual mechanism removed, named \textsc{No-DTeachers}, \textsc{No-DStudents}, and \textsc{No-Cond}, respectively. Specifically, \textsc{No-DTeachers} means that no voice conversion module is introduced; \textsc{No-DStudents} means that no transformation or contrastive loss is imposed in the student module; \textsc{No-Cond} means that no speaker embeddings are fed to the predictor.  Their respective performance is reported in Table~\ref{tab:ablation}. As can be seen, all three models perform significantly worse than \algnamens. We can conclude that all three modules are essential to \algnamens.

\begin{figure}[!t]
     \centering
     \begin{subfigure}{0.7 \columnwidth}
         \centering
         \includegraphics[width=\columnwidth]{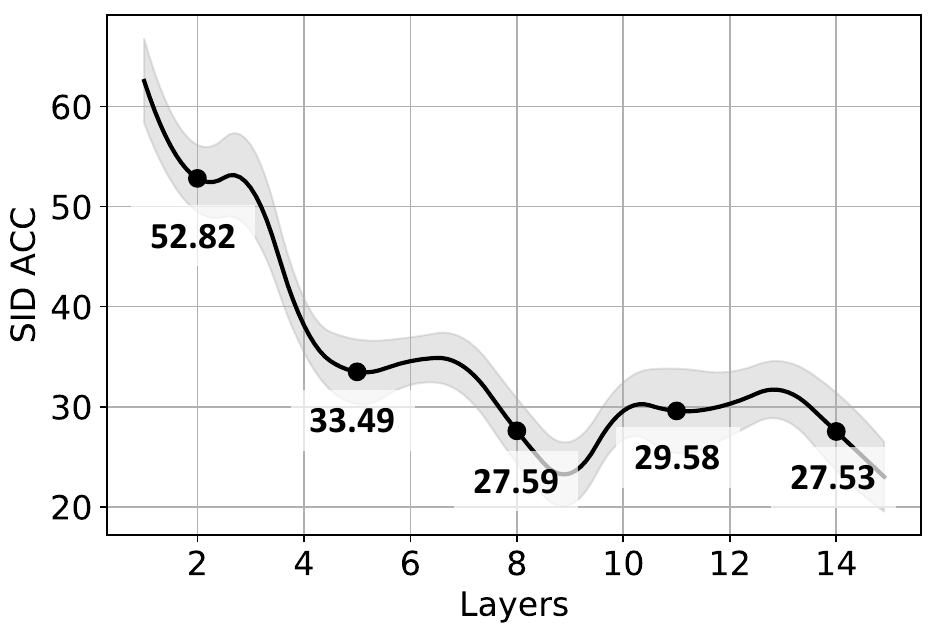}
         \caption{The \algname model}
         \label{fig:flow1}
     \end{subfigure}
     \hfill
     \begin{subfigure}{\columnwidth}
         \centering
         \includegraphics[width=0.7\columnwidth]{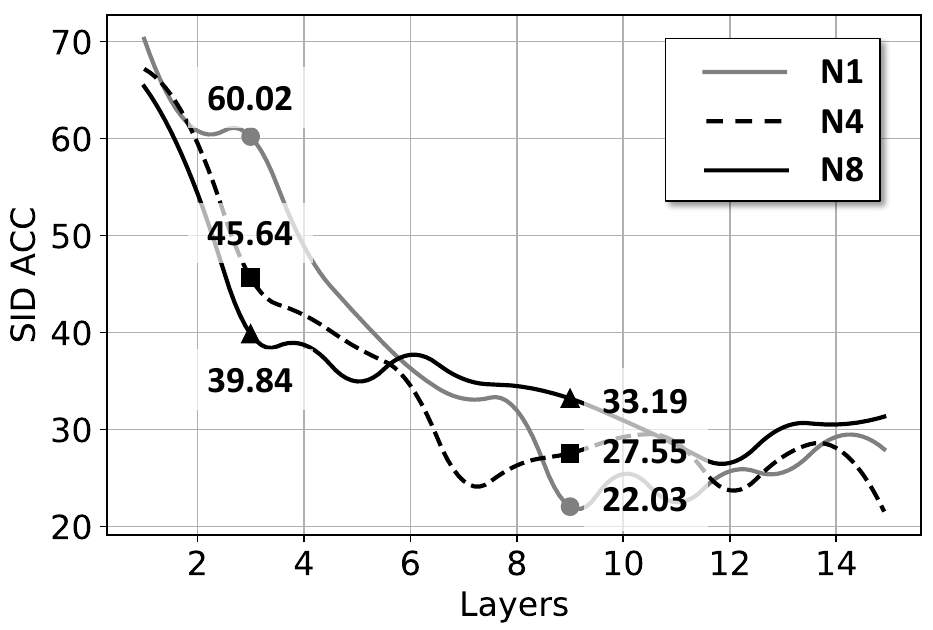}
         \caption{Comparison of \algnamens-N1, \algnamens-N4, and \algnamens-N8}
         \label{fig:flow2}
     \end{subfigure}
    \caption{SID accuracies as functions of network layers}
    \label{fig:flow}
    \vspace{-0.2in}
\end{figure}

\paragraph{Speaker Information Flow} In order to verify if our conceptual curve in Figure~\ref{fig:concept} is correct, we evaluate the SID accuracy (as a proxy of speaker information) on \emph{every layer} of \algnamens. For each experiment, we select the best SID model based on validation accuracy for every $10$k iterations and compute the accuracy achieved by all these SID models to gauge statistical significance. Figure~\ref{fig:flow}(a) plots the SID accuracy as a function of network layers. By comparing Figures~\ref{fig:concept} and \ref{fig:flow}(a), we find that our key hypotheses are verified. First, the SID monotonically descreases before layer 9, and there is a significance decrease before layer seven, which is roughly where the contrastive loss is imposed.\footnote{The contrastive loss is always imposed at the last but fifth layer. Due to layer drop, the actual layer at which the contrastive loss is imposed is randomly distributed before layer seven, with the major probability mass in layers five to seven.} This observation is consistent with the sharp decrease in Figure~\ref{fig:concept}. Second, there is a slight increase towards the last layers, which is consistent with the increase in Figure~\ref{fig:concept}, although it occurs at the lower level than we expect. A likely reason is that our SID is performed on the discrete representations, whereas our conceptual curve is based on continuous representation. Nevertheless, these observations confirm the rationales behind our model design.

\paragraph{Position of Contrastive Loss} To illustrate how the position of contrastive loss influences the speaker information flow, we run three variants of \algname, denoted as \algnamens-N1, \algnamens-N4, \algnamens-N8. ``N1'', ``N4'' and ``N8'' mean that the contrastive loss is imposed in last, last but three, and last but seven layers, respectively (instead of last but five as in our standard \algnamens). Figure~\ref{fig:flow} compares the SID accuracy against the layers for these variants. We can observe a similar decreasing and then increasing trend in all these curves. More importantly, there is an interesting correlation between the SID accuracy drop positions and the layer at which contrastive loss is imposed. \algnamens-N8 drops earliest, followed by \algnamens-N8 and then \algnamens-N1. We can also observe that the model that starts dropping late would drop to a \emph{lower} point. 

We also evaluate the performance of these model variants with respect to the aforementioned metrics, and the results are listed in Table~\ref{tab:ablation}. As can be observed, all the variants achieve very competitive results, outperforming all the baselines. 
These results show that the competitiveness of \algname does not hinge on specific choices of this hyperparameter.
\vspace{-0.1in}
\paragraph{Contrastive Loss Weight} We also evaluate different contrastive loss weights (the $\lambda$ in Equation~\eqref{eq:loss_combine}). Recall that our standard \algname uses a weight coefficient of 1e-5. The following four weight coefficient are tested, 1e-6, 5e-6, 2e-5, and 5e-5, and the results are listed in Table~\ref{tab:ablation}. We can observe that \algname performs competitively across all the models, showing a low hyperparameter sensitivity.

\section{Conclusions}
In this paper, we propose \algnamens, which is a speech representation learning network that aims to remove speaker information while preventing loss of content information. \algname builds upon the \textsc{HuBERT} framework and introduces three key disentanglement components: disentanglement in teachers, disentanglement in students, and teacher conditioning of the predictor. Our empirical analyses confirm that all three modules are essential to the success of \algnamens. We also verified that a successful speaker disentanglement does help with a wide range of content-related speech processing tasks.
Meanwhile, \algname still have some limitations, \emph{e.g.} slight content loss and the lack of hyperparameter selection method.
Finding solutions to these problems will be our future directions.

\bibliography{example_paper}
\bibliographystyle{icml2022}

\newpage
\clearpage
\appendix
\section{Additional Experiment Results}
\label{app:generation}

\subsection{Visualizing Speech Generation Cluster Sequences}
We compare \textsc{HuBERT-iter} and \algname by visualizing their speech labels on a small subset of \texttt{Librispeech} and calculating the average distance between male and female speakers speaking the same content.

\begin{figure*}
    \centering
    
    \includegraphics[trim={7cm 2.1cm 6cm 0}, ,clip, width=2\columnwidth]{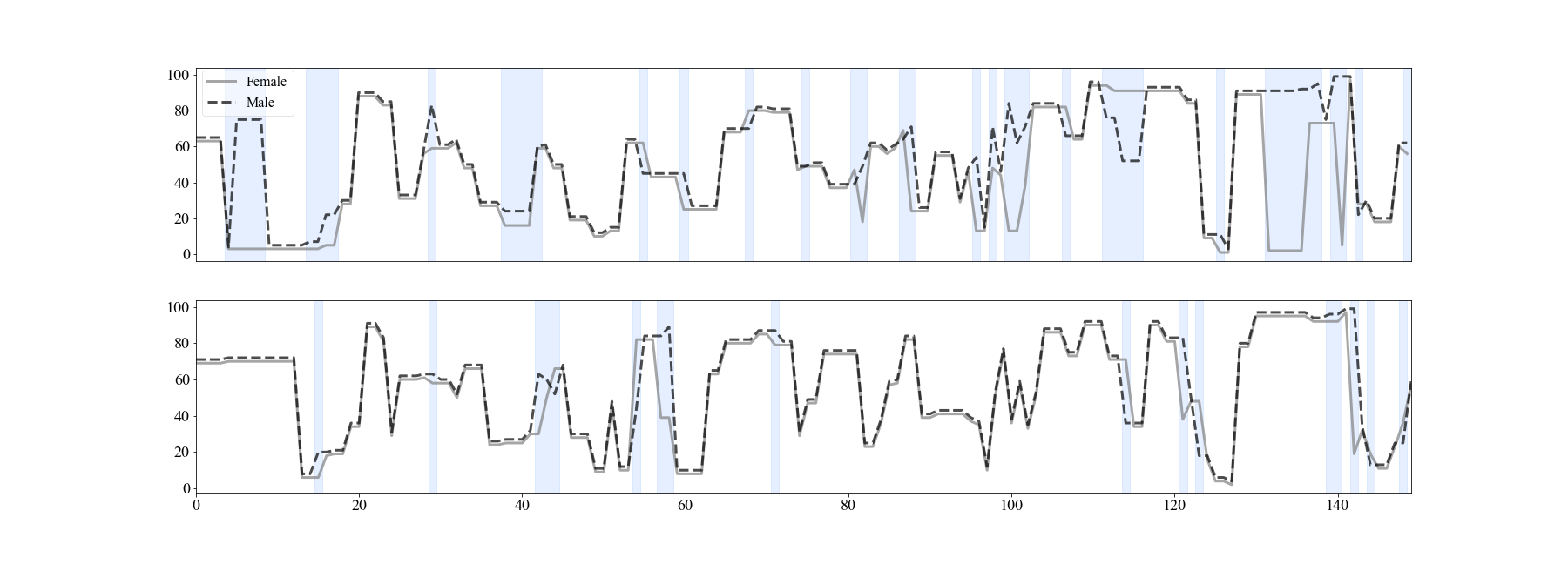}

    \caption{Example label rank sequences generated using \textsc{HuBERT-iter} (upper) and  \textsc{ContentVec} (lower). The mismathed frames are highlighted using blue.}
    \label{fig:label_rank}
\end{figure*}

For each $k$-means label, we compute its ratio of occurrence between male and female speakers in the \texttt{dev-clean} split of \texttt{Librispeech}. We then rank the labels by the occurrence ratio of female; the more frequently the label occurs in female utterances, the larger rank it gets. We randomly selected 100 female utterances from \texttt{train-clean-100} split of \texttt{Librispeech} and convert them to male speeches using the voice converter. Then for each pair of male and female utterances with the same content, we map their label sequences to ranked label sequences. Since each pair of utterance shares the same content, their corresponding ranked label should ideally be the same. 
Figure \ref{fig:label_rank} shows the ranked label sequences on an example male-female pair, where their differences are highlighted with blue blocks.  We observe that there are fewer blue blocks in \algname plot and therefore the male and female curves align better using \algname labels. This observation implies that \algname disentangles more speaker information from the content compared to \textsc{HuBERT-iter}.

To confirm this observation, we apply additional dynamic time warping to align the utterance pairs and then compute the average L0 distance in their ranked labels sequences over frames. Dynamic time warping is necessary as we find that the converted audio still have slight mismatch in speech rate. The result is shown in Table \ref{tab:l0}; \algname does have lower average L0 distance, indicating the gender information is hardly preserved in the \algname features.

\begin{table}[h]
\caption{Results of average L0 distance between male and female speech pairs}
\label{tab:l0}
\vskip 0.15in
\begin{center}
\begin{small}
\begin{tabular}{l|c|c}
\toprule
&   \textsc{HuBERT-iter} & \algname    \\
\midrule
L0 Distance  & 0.165 & \bf 0.118 \\
\bottomrule
\end{tabular}
\end{small}
\end{center}
\vskip -0.1in
\end{table}

\subsection{ASR Transcriptions of Example Generated Speech}
In Table \ref{tab:example} we show several example transcripts of speech generated by \textsc{HuBERT-iter} and \algname, using the first three seconds of utterances from the \texttt{test-clean} split of the \texttt{Librispeech} dataset.

It seems the speech representation from \algname is better at reconstructing the prompts than that from the \textsc{HuBERT-iter}. Both models are able to generate meaningful speech within a few words after the prompt. However, neither of them is capable of generating meaningful long sentences. 

\begin{table}[h]
\caption{Example transcripts of resynthesized speech using prompt. \textit{H} denotes the transcripts from \textsc{HuBERT-iter} and \textit{C} denotes those from \algname. The highlighted texts are resynthesized from the given prompt and the non-highlighted texts are from labels generated by language models}
\label{tab:example}
\vskip 0.15in
\begin{center}
\begin{small}
\begin{tabular}{l|cccc}
\toprule
Prompt & \multicolumn{4}{l}{after early nightfall the yellow lamps} \\\midrule
\multicolumn{5}{p{8cm}}{\sethlcolor{light-gray}  
    \textit{H}: \hl{after early nightfall the yellow lamps} would light i double loud in my hands and a couple of dances told me we
}  \\
\multicolumn{5}{p{8cm}}{\sethlcolor{light-gray}  
    \textit H: \hl{ah after early nightfall meat you lamps} would light in that six thous heaven when happens
}  \\\midrule

\multicolumn{5}{p{8cm}}{\sethlcolor{light-gray}  
    \textit C: \hl{after early nightfall the yellow lamps} would light and they were upon their     liberty until morning when they were} \\
\multicolumn{5}{p{8cm}}{\sethlcolor{light-gray}  
    \textit C: \hl{after early nightfall the yellow lamps} would light and chill in some dim spectacle lancers in the sun
}\\\midrule\midrule


Prompt&\multicolumn{4}{l}{if a layman in giving baptism pour the water} \\\midrule
\multicolumn{5}{p{8cm}}{\sethlcolor{light-gray}  
    \textit H: \hl{if alane and giving baptism for the waters} which dry even stream formed the foreheads of
}  \\
\multicolumn{5}{p{8cm}}{\sethlcolor{light-gray} 
    \textit H: \hl{fish iffen men and giving baptism for the water} of flame and he came on a great
}  \\\midrule

\multicolumn{5}{p{8cm}}{\sethlcolor{light-gray}  
    \textit C: \hl{if a layman in giving baptism pour the water} forth to the knot and come out without any
} \\
\multicolumn{5}{p{8cm}}{\sethlcolor{light-gray}  
    \textit C:\hl{if the laymen in giving baptism pour the water} in a shower of sparkle which made these sure
} \\
\bottomrule
\end{tabular}
\end{small}
\end{center}
\vskip -0.1in
\end{table}

\subsection{Voice Converter Quality}

Since the voice converter in teacher is a key mechanism of \algnamens, we would like to investigate how the quality of the voice converter impact the quality of \algnamens. To this end, we performed an ablation study where we replaced the voice converter with a compromised version trained with fewer number of steps (10,000 steps instead of 40,000 steps). Accordingly, the best-layer ABX(w), ABX(a) and PNMI degrade to 6.05, 7.78, and 0.5616 respectively (The best layer for these three metrics are 8, 12, 9, respectively). This result shows that the quality of the voice converter is very essential to the performance of \algnamens. A poor voice converter would produce an even worse performance than the variant with no voice converter at all, as indicated by the `NO-DT' results in Table~\ref{tab:ablation}.

\subsection{Contrastive Loss on Multiple Layers}

So far, we have only evaluated on \algname models where the contrastive loss is imposed on only one layer. To investigate how the performance will change if the contrastive loss is imposed on multiple layers, we perform an ablation study where the contrastive loss is simultaneously imposed on the last, last but third, and last but seventh layers. To keep the scale consistent, the weight on each contrastive loss term is reduced to 1/3 of its original value. The best-layer ABX(w), ABX(a) and PNMI become 5.19, 6.53, and 0.5684 respectively (The best layer for these three metrics are all 12), which is slightly worse than the original \algname model. In other words, there is no evidence that imposing contrastive loss on multiple layers can improve performance of \algnamens.

\section{Experiment Details}

\subsection{\algname Implementation and Training}
\label{app:implementation_training}




The speech representation network of \algname has the same architecture as the \textsc{HuBERT}, which has 7 temporal convolutional feature extraction blocks followed by 12 layers of transformer layers of model dimension 768. During training, each layer is independently dropped with a probability of 0.05. According to Section~\ref{subsec:flow}, the speaker-disentanglement is optimal at the output of the speech representation network, so we select the output-layer representation as the \algname features. The contrastive loss is imposed at the last but five layer, the temperature $k$ is set to 0.1, and the contrastive loss weight $\lambda=\text{1e-5} * num\_train\_steps$, which linearly increases to 10 when training for 100k steps. 

The masking strategy is the same as in Wav2Vec 2.0 \cite{Baevski2020wav2vec2A}, with the masking probability set to 0.08. The masks of the two transformation paths are the same. It is also worth mentioning that the masking is only imposed for the masked prediction loss, \e{\mathcal{L}_{pred}}. No masking is imposed for the contrastive loss, \e{\mathcal{L}_{cont}}.

The predictor of \algname contains three transformer layers \emph{without} layer drop. Each transformer layer uses the conditional layer normalization to inject speaker information, of which the scale and bias are learnable linear mappings of the conditioned speaker embedding. The frame masking scheme and prediction logit generation for the masked prediction task are the same as \textsc{HuBERT}.

Our model is trained for 100k steps using 36 GPUs, with a batch size of at most 76 seconds of audio per GPU, which takes about 19 hours to complete. The best model is selected based on the lowest validation masked prediction loss. 

To generate the teacher labels, we use the voice converter proposed by \citet{polyak2021speech}, which is trained by reconstructing speech from quantized speech representations. We directly use the publicly available implementation\footnote{\url{https://github.com/facebookresearch/speech-resynthesis}}, and all hyperparameters are kept the same unless explicitly mentioned. We re-train the resynthesis model on a subset of 200 speakers chosen from the \texttt{Librispeech} dataset \cite{Panayotov2015LibrispeechAA}. Of the 200 speakers, 100 of them are from the \texttt{train-clean-100} or the \texttt{train-clean-360} subset, while the other 100 speakers are from the \texttt{train-other-500} subset. The ratio of male and female speakers is 1:1. These speakers are chosen based on the amount of audio available as listed in the metadata. The input consists of three components: the output from a speech-to-unit model, the output from a pitch-to-unit model, and speaker embedding. We choose the publicly-released \textsc{HuBERT-Base} model\footnote{\url{https://github.com/pytorch/fairseq/tree/main/examples/hubert}} as the speech-to-unit model. To extract the discrete speech units, we train a k-means model with 100 clusters on the features extracted from the seventh layer of the \textsc{HuBERT-Base} model, using the \texttt{train-clean-100} subset. The pitch-to-unit model is a VQ-VAE \cite{Oord2017NeuralDR} with a convolutional encoder plus a bottleneck, trained to reconstruct per-speaker mean-variance normalized pitch contours. The speaker embedding is based on d-vectors \cite{Heigold2016EndtoendTS}, and we use an existing implementation\footnote{\url{https://github.com/resemble-ai/Resemblyzer}}. To resynthesize speech, the inputs are fed into a decoder based on \texttt{HiFi-GAN} \cite{Kong2020HiFiGANGA}, except that the input format is modified. The decoder is trained for a total of 460k steps, and each step consists of alternating updates between the generator and the discriminators. After training, we randomly chose eight unseen speakers from the \texttt{dev-clean} and the \texttt{dev-other} subsets, with an equal number of male and female speakers. For each saved checkpoint, we convert the utterances from those eight unseen speaker to the voices of ten seen speakers from the 100 clean speakers used for training. The conversion is done simply by changing the speaker embedding to those of the target speakers. For each target speaker, the speaker similarity is calculated as the cosine similarity between the speaker embedding obtained with the converted utterances, and the ground-truth target speaker embedding extracted using the target speaker's own utterances. The saved iteration that gives the highest average target speaker similarity on the ten seen speakers is selected. After that, we further select a target speaker for the teacher utterances from the 100 clean speakers seen during training. This is obtained by converting the utterances from the unseen speakers into each of the 100 clean speakers with the selected model iteration, and then calculating the same cosine similarities as stated above. The teacher utterances are then generated by converting the entire \texttt{Librispeech-960} to the voice of that selected target speaker (which is a male speaker), with the selected model iteration.

\subsection{Baselines and Dataset}
\label{app:baselines_dataset}



$\bullet$ \textsc{Wav2Vec 2.0} \cite{Baevski2020wav2vec2A}: Following \citet{Lakhotia2021OnGS}, the 14th layer representation is chosen.

$\bullet$ \textsc{HuBERT} \cite{Hsu2021HuBERTSS}: We adopt the publicly-released pretrained model in \citet{Ott2019fairseqAF}. Following \citet{Hsu2021HuBERTSS}, the sixth layer representation is chosen.

$\bullet$ \textsc{HuBERT-iter}: Since \algname is guided by a pretrained \textsc{HuBERT} as teachers, for a fair comparison, we introduce another \textsc{HuBERT} baseline that is trained by the same pretrained \textsc{HuBERT} as teachers (except that no voice conversion is performed). This baseline controls the benefit of iterative training. We performed the same teach quality evaluation as in \citet{Hsu2021HuBERTSS}, and identified layer eight as the best performing layer.

\algname and all the baselines are trained on the full 960 hours of \texttt{Librispeech} dataset \citep{Panayotov2015LibrispeechAA}. The teacher quality evaluations are conducted on the dev-clean and dev-other partition of \texttt{Librispeech}, and we use publicly available phone alignment using the Montreal Aligner. \footnote{\url{https://https://github.com/CorentinJ/librispeech-alignments}}  If the evaluation task requires discrete representations, all the representations will be quantized to 100 clusters by $k$-means except for SUPERB, which uses 500 clusters. Otherwise, the continuous representations will be used.

\subsection{Zero-shot Content Probe}

We evaluate our models on set of zero-shot probing tasks proposed in the Zero-Resource Speech Challenges \cite{Rivire2021TowardsUL, Dunbar2021TheZR}, because they require a high alignment between the discrete representation and phonetic content. For some of these tasks, a language model trained on the discrete speech representations is needed. We use the same language model and hyperparameter setting as in \citet{Lakhotia2021OnGS}, which is the \emph{transformer LM big architecture} implemented in \texttt{fairseq} \citep{Ott2019fairseqAF}. This transformer model has 12 layers with 16 attention heads, embedding size of 1024, FFN size of 4096 and dropout probability of 0.1. The model is trained on eight 32-GB GPs for 20000 updates the using distributed training with a dynamic batch size that contains at most 4096 tokens. The learning rate is set to $5\times10^{-4}$. We select the model that has the lowest loss on the validation set.
In this experiment, we also test the models with 200 clusters.








\begin{table*}[!t]
\caption{Best layers of the ablation models}
\label{tab:best_layer}
\begin{center}
\begin{small}
\begin{tabular}{l|ccc|ccc|cccc}
\toprule
& \multicolumn{3}{|c|}{Removing each module}  & \multicolumn{3}{|c|}{Position of $\mathcal{L}_{cont}$} & \multicolumn{4}{|c}{Weight of $\mathcal{L}_{cont}$ ($\lambda$)} \\
 & \textsc{No-DT} & \textsc{No-DS} & \textsc{No-Cond} & \textsc{N1} & \textsc{N4} & \textsc{N8} & 1e-6 & 5e-6 & 2e-5 & 5e-5\\
\midrule
ABX(w)  & 12 & 12 & 10 & 12 & 12 & 12 & 12 & 12 & 12 & 12 \\
ABX(a)  & 12 & 12 & 10 & 12 & 12 & 12 & 12 & 12 & 12 & 12 \\
PNMI  & 12 & 11 & 11 & 12 & 12 & 12 & 12 & 12 & 12 & 12 \\
\bottomrule
\end{tabular}
\end{small}
\end{center}
\end{table*}

\subsection{Language Modeling}

Language models built directly on the discrete speech representations can be applied to many content-related speech generation and analysis tasks. We would like to explore whether a disentangled speech representation can contribute to a language model with higher quality. To this end, we use the same language models as in Section~\ref{subsec:zero-speech} to generate random speech representation sequences under different temperatures, resynthesize speech from these sequences and transcribe the audio using a pretrained speech recognizer. 

We adopt the same \textsc{Tacotron}-based speech synthesizer as in \citet{Lakhotia2021OnGS}, and finetune it on \texttt{LJ Speech} dataset \citep{ljspeech17} that contains only one female voice. The model is trained on eight 32-GB GPUs with a batch size of 32 for 20000 iterations. The learning rate is set to 0.001. We use the model from the last iteration for speech resynthesis.

We use \textsc{Wav2Vec2.0-Large} model pretrained on the \texttt{Libri-Light} dataset without finetuning to generate transcriptions. Note that the \textsc{Tacotron} model was pretrained on 22050 Hz speech and works best with 22050 Hz. We need to resample the generated speech to 16000 Hz so that \textsc{Wav2vec} can work correctly.

For each temperature level, we sample speech sequences from language model, resynthesize audio using these sequences. We compute the perplexity (PPX) and VERT score of the transcript. VERT score is the geometric mean of the self- and auto-BLEU as proposed in \citet{Lakhotia2021OnGS}. 

The perplexity is evaluated using a pretrained English language model \emph{transformer\_lm.wmt19.en} from \texttt{fairseq}.
BLEU scores measures the similarity between two sentences and self-BLEU score \citep{10.1145/3209978.3210080} measures the similarity between sentences generated using the same prompt. Higher self-BLEU score means less diversity of the generated speech.

Auto-BLEU score \citep{Lakhotia2021OnGS} measures within-sentence diversity and is computed as the ratio of $k$-grams that are repeated at least once:
\begin{equation}
    \small
    \texttt{AUTO-BLEU}(u,k)=\frac{\sum_s \mathbb{I}[s\in(\text{NG}_k(u)\setminus s) ]}{|\text{NG}_k(u)|},
    \label{eq:loss_auto_bleu}
\end{equation}
where $\text{NG}_k(n)$ is the set of $k$-grams of utterance $u$.

All the perplexity-VERT pairs at different temperature levels form a curve depicting the quality-variety trade-off of each language model. We use the perplexity and VERT scores of the ground-truth text as an anchor point to compute AUC: the area above the anchor point and under the perplexity-VERT curve. We use AUC as the tradeoff measure; lower AUC means the model is closer to the anchor point, therefore closer to ground-truth text.






\subsection{Voice Conversion}
For the voice conversion experiments, we use the same voice conversion model proposed by \citet{polyak2021speech} and follow the same re-synthesis training described in Appendix \ref{app:implementation_training}, except for the input, the input layer and train/dev/test split. In order to compare how well our \algname system disentangles speaker information with the two baseline models (\textsc{HuBERT} and \textsc{HuBERT-iter}), we extract the continuous speech unit of the best performing layer of all three models (sixth layer for \textsc{HuBERT}, eighth layer for \textsc{HuBERT-iter} and \algnamens). 

The 200-speaker sub-corpus created from \texttt{LibriSpeech} in Appendix \ref{app:implementation_training} is further splitted so that around 70\% of the utterances for each speaker serve as the training set, around 10\% serve as the development set, and around 20\% serve as the test set. The decoder is trained for a total of 440k steps, and we choose the best performing checkpoint (over the last 200k steps) by re-synthesizing the development set and then averaging the per-speaker cosine similarities. The per-speaker cosine similarity is calculated as dot-product between the length-normalized d-vector \citep{Heigold2016EndtoendTS} extracted from re-synthesized development-set utterances of that speaker and the speaker's original (length-normalized) d-vector used in training. Voice conversion is carried out by choosing 10 seen speakers that belong to the \texttt{train-clean-100} or \texttt{train-clean-360} subsets of \texttt{LibriSpeech} and another 10 seen speakers that belong to the \texttt{train-clean-other} subset. All test utterances are converted to the 20 target speakers (the utterances from those 20 speakers are only converted to the other 19 target speakers). We separate the test set into a `clean' subset and an `other' subset, and we separately calculate the average-speaker cosine similarity for the 10 `clean' target speakers and the 10 `other' target speakers. For each of the four scenarios, the average cosine similarity is obtained by averaging the per-speaker cosine similarities over the 10 target speakers in that scenario. The per-speaker cosine similarity is calculated as the dot product between a length-normalized d-vector obtained from all the converted speech of a specific target speaker and that speaker's original d-vector used in training.







\subsection{Ablation Studies}



For each metric, we report the best-layer result. The best layers of each ablation model and each metric shown in Table~\ref{tab:best_layer}.


\end{document}